\documentclass[authoryear]{elsarticle}
\usepackage[colorlinks=true,linkcolor=black, citecolor=blue, urlcolor=blue]{hyperref}

\usepackage{amsmath}
\usepackage{amssymb}
\usepackage{graphicx,psfrag,epsf}
\usepackage{enumerate}
\usepackage{natbib}
\usepackage{multirow}
\usepackage{longtable}
\usepackage{booktabs}
\usepackage{xcolor}
\usepackage{url} 

\newtheorem{Lem}{Lemma}

\journal{Econometrics and Statistics}

\begin{document}

\begin{frontmatter}



\title{A Unified Frequency Domain Cross-Validatory Approach to HAC Standard Error Estimation}

\author[inst1]{Zhihao Xu\corref{cor1}}

\affiliation[inst1]{organization={Department of Statistics and Data science, Yale University},
            country={USA}}

\author[inst2]{Clifford M. Hurvich}

\affiliation[inst2]{organization={Stern School of Business, New York University},
            country={USA}}

\cortext[cor1]{Corresponding author. Email: zhihao.tonyxu@gmail.com.\\
\indent {The supplementary material contains the R code to perform REML estimation of autoregressive models using the PCG method described in the article.}}

\begin{abstract}

A unified frequency domain cross-validation (FDCV) method is proposed to obtain a heteroskedasticity and autocorrelation consistent (HAC) standard error. This method enables model/tuning parameter selection across both parametric and nonparametric spectral estimators simultaneously. The candidate class for this approach consists of restricted maximum likelihood-based (REML) autoregressive spectral estimators and lag-weights estimators with the Parzen kernel. Additionally, an efficient technique for computing the REML estimators of autoregressive models is provided. Through simulations, the reliability of the FDCV method is demonstrated, comparing favorably with popular HAC estimators such as Andrews-Monahan and Newey-West.

\end{abstract}


\end{frontmatter}


\section{Introduction}
\label{sec:intro}

Many regression problems involving economic and financial time series suffer from autocorrelated errors. Under such circumstances, the standard error estimator for the OLS coefficients derived under the uncorrelated errors framework will no longer be consistent. Hence, the confidence interval and test statistics that are based on the usual standard error estimator of the OLS coefficients cannot be relied upon, even in large samples. In the past 30 years, several heteroskedasticity and autocorrelation consistent (HAC) standard error estimators have been proposed, most notably by \cite{newey1987simple}, \cite{newey1994automatic}, \cite{andrews1991heteroskedasticity}, and \cite{andrews1992improved}. These methods are implemented in widely-used statistical packages. 

The focus of this paper is on the special case of estimating the standard error of the sample mean of a stationary univariate time series. Discussion of extensions to a general time series regression setting is deferred to Section 6. Thus, the location model is considered: $$X_t = \mu + \varepsilon_t, \qquad  t=0,\cdots, n-1,$$
where $\mu$ is the unknown population mean and $\{\varepsilon_t\}$ is a zero-mean weakly stationary time series. The ordinary least-squares (OLS) estimator of $\mu$ is the sample mean 
$$\hat{\mu} = \overline{X_n} = \frac{1}{n}\sum_{t=0}^{n-1}X_t.$$
Under the assumption that $\{X_t\}$ is weakly stationary and has short memory (the spectral density of $\{X_t\}$ at zero frequency $f(0)$ is finite and positive), the aim is to conduct valid inference for $\mu$ when the $X_t$ are autocorrelated.
Let $c_j = Cov(X_t,X_{t-j})$ be the lag-$j$ autocovariance sequence. It is easily shown that  $$Var(\hat{\mu}) =  \frac{1}{n}\Big(c_0+2\sum_{j=1}^{n-1}\frac{n-j}{n}c_j\Big).$$
Furthermore, under suitable regularity conditions,
$$c_0+2\sum_{j=1}^{n-1}\frac{n-j}{n}c_j \xrightarrow{n \to \infty}\sum_{j = -\infty}^{\infty} c_j.$$
The term $S^2 = \sum_{j = -\infty}^{\infty} c_j$ is referred to as the long-run variance. The long-run variance is closely connected with the spectral density of $\{X_t\}$, assumed here to exist, and to be given by $$f(\omega) = \frac{1}{2\pi} \sum_{j = -\infty}^{\infty}c_j \exp(i\omega j), \qquad \omega \in [-\pi,\pi].$$
Specifically, it can be shown that
$$n Var(\overline{X_n}) \rightarrow S^2 = 2\pi f(0).$$

Estimation of the spectral density at zero frequency has been at the heart of the HAC problem of the past 30 years and is particularly relevant for univariate tests such as certain unit root tests (e.g., \cite{phillips1988testing} ) and stationarity tests (e.g., \cite{kwiatkowski1992testing}), as well as multivariate applications like hypothesis testing in regression models. Spectral density is a frequency domain concept, and an overview of concepts and techniques of frequency domain time series analysis is provided in Section 2. One can estimate $f(0)$ parametrically or nonparametrically. This paper primarily focuses on the univariate case and develops the FDCV method for HAC estimation in this context. Although potential extensions to the multivariate case are briefly discussed in the conclusion, this paper does not provide a comprehensive solution for the multivariate setting. Nonetheless, this paper aims to explore the potential of FDCV in the context of univariate HAC estimation. To estimate $f(0)$ parametrically, one can estimate a time series model and construct the value of $f(0)$ that this model implies. One of the most popular parametric estimates is the autoregressive spectrum estimate. In practice, the user needs to choose the order of the autoregressive model. On the other hand, one can estimate $f(0)$ nonparametrically. A well-known class of nonparametric estimators are the lag-weights estimators. The tuning-parameter selection problem in this case is the choice of the truncation point or bandwidth (as well as, potentially, the choice of the kernel function). 

The first-generation solutions to the HAC problem are based on the goal of minimizing some local criterion such as mean squared error (MSE) of the estimator $\hat{f}(0)$
$$MSE(\hat{f}(0)) = E\Big[\Big(\hat{f}(0)-f(0)\Big)^2\Big],$$
using a lag-weights estimator (See \cite{andrews1991heteroskedasticity}).
The choice of the kernel can be motivated by the asymptotic properties of the corresponding estimate, but the choice of the truncation point is much more challenging since the optimal choice in terms of criteria like MSE depends on the actual unknown spectral density. In practice, \cite{andrews1991heteroskedasticity} and \cite{andrews1992improved} propose plug-in approaches to estimate the optimal truncation point or bandwidth. Unfortunately, in small to moderate sample sizes, the first-generation HAC estimator based on nonparametric estimation has a substantial mean squared error under certain data-generating mechanisms. In particular, the nonparametric estimator will have desirable performance if the spectral density function is relatively flat around zero, such as the case of white noise but will have a substantial bias if the spectrum has a strong peak at zero frequency. 

The second-generation solution to the HAC problem uses the idea of prewhitening to address the bias issue. \cite{andrews1992improved} propose using a fixed-order autoregressive filter to transform the data such that the spectrum of the transformed data will be flatter in a neighborhood of zero frequency and therefore the nonparametric estimator will be less biased. The idea of prewhitening was subsequently implemented by \cite{andrews1992improved} and \cite{christiano1996small} as a part of their HAC estimators. The HAC literature measures performance by the coverage rates of the confidence intervals for the regression parameters. In the simulation study of \cite{andrews1992improved}, such fixed-order prewhitening can improve the performance of the coverage probability in many cases. In Andrews and Monahan's simulation study, the filter is an $AR(1)$ filter based on the least-squares estimator of the autoregressive model.

What is considered as the second-generation answer to the HAC problem is an attempt to combine the parametric approach and nonparametric approach. The fixed-order autoregressive filter serves as a parametric component and the nonparametric estimator allows for the non-flatness in the spectral density of the prewhintened data. However, as these approaches are currently implemented, there is no model selection of the order of the prewhitening filter. \cite{den199712} perform a simulation study that shows the drawbacks of fixed-order prewhitening. They show that if the first-order autocorrelation of the prewhitened series is small, but higher-order autocorrelation coefficients are substantial, the confidence interval of prewhitening-based HAC methods tend to significantly overcover or undercover $\mu$. They point out that such poor performance is due to fixed-order prewhitening. Furthermore, using least-squares to estimate the autoregressive filter may not be desirable. If the data generating process has a strong peak in the spectral density at or near zero frequency, using an $AR(1)$ filter based on the least-squares estimator fails in prewhitening the data. In this case, the spectral density of the transformed data still has a substantial peak around zero frequency. But if the underlying data generating process is really an autoregression, then a good parameter estimator, such as the restricted maximum likelihood (REML) estimator (See \cite{cheang2000bias}, \cite{harville1974bayesian} and \cite{chen2012restricted}), will lead to a good prewhitening. If it were known that the actual data generating mechanism was truly an autoregression, then the parametric autoregressive spectral density estimator should be used rather than the nonparametric approach. The central issue, however, is that the actual data generating mechanism is not known. To address the issue of not knowing the actual data generating mechanism, this paper introduces a method allowing for unified model / tuning parameter selection across both parametric and nonparametric estimators. The introduced unified model / tuning parameter selection for HAC standard error estimation is based on the idea of frequency domain cross-validation (FDCV).

FDCV was originally proposed by \cite{wahba1975periodic} to select the tuning parameter of spline-based nonparametric spectrum estimates. \cite{beltrato1987determining} propose a cross-validated log likelihood (CVLL) for cross-validation in the frequency domain to select the bandwidth of average periodogram estimates. \cite{hurvich1985data} uses the cross-validation function of \cite{wahba1975periodic}, but instead of restricting attention to
splines, he allows for an arbitrary estimator of the spectral density. Hurvich defines a frequency domain leave-out-one version of any spectrum estimate, opening up the possibility for unified selection among several types of estimators, simultaneously including nonparametric estimators and parametric estimators. All of the frequency domain cross-validation methods described above originally focused on the entire frequency range, $[0,\pi]$. The use of this global frequency band makes such methods apparently incompatible with the problem of HAC as HAC focuses on the spectrum at zero frequency.

In this paper, a localized version of FDCV for the HAC problem is proposed, based on a class of candidates that simultaneously includes both autoregressive (REML-based) estimates and nonparametric estimates. It is also shown how to compute the REML estimator with the preconditioned conjugate gradient (PCG) algorithm and demonstrated that this allows the evaluation of the restricted likelihood function in $O(n\log n)$ operations. Simulations will be examined to compare the coverage rates of the resulting confidence interval for $\mu$ with the coverage rates corresponding to the Newey-West and Andrews-Monahan methods. Section 2 provides an overview on FDCV and spectrum estimation. Section 3 introduces a unified approach based upon FDCV for HAC standard error estimation. Section 4 presents an efficient method of computing the REML-based AR parameter estimates. Section 5 reports Monte-Carlo results for several kernel-based HAC methods and FDCV. Section 6 provides some concluding remarks.

\section{An Overview of FDCV and Spectrum Estimation}
\label{sec:overview}
Let $\{x_t\}_{t=0}^{n-1}$ be a real-valued data set. The discrete Fourier transform (DFT) of $\{x_t\}_{t=0}^{n-1}$ is defined as the sequence of complex numbers $$J_j = \frac{1}{n}\sum_{t=0}^{n-1}x_t \exp(-i\omega_jt), \quad j=0,...,n-1,$$
where $\omega_j$ is the $j$-th Fourier frequency defined as $\omega_j = \frac{2\pi j}{n}$.\\
The sequence $\{x_t\}$ can be retrieved from the DFT sequence $\{J_j\}_{j=0}^{n-1}$ using the inverse Fourier transform
$$x_t =\sum_{j=0}^{n-1}J_j\exp(i\omega_j t), \qquad t=0,...,n-1.$$
The periodogram $I(\omega_j)$ at Fourier frequency $\omega_j$ is defined as $$I(\omega_j) = \frac{n}{2\pi}|J_j|^2.$$

Given a zero-mean weakly stationary time series $\{X_t\}$, a widely-used estimate of $c_r$ for all $r$ such that $|r|<n$ is the sample autocovarinance $\hat{c}_r = \frac{1}{n}\sum_{t = |r|}^{n-1} x_tx_{t-|r|}$. The periodgram can be expressed in terms of $\{\hat{c}_r\}$ as
$$I(\omega_j) = \frac{1}{2\pi}\sum_{|r|<n}\hat{c}_r \exp(ir\omega_j).$$
A widely-used nonparametric spectrum estimator in both the spectral density estimation literature and the HAC literature is the lag-weights (also called Blackman-Tukey) estimate, defined as 
$$\hat{f}(\omega) = \sum_{|r| \leq h} w (\frac{r}{h}) \hat{c}_r \exp(ir\omega),$$
where $h$ is a non-negative integer, called the truncation point. The function $w(x)$ is called the lag window or kernel. There are two critical questions regarding the lag-weights estimate: the choice of kernel and the choice of truncation point $h$. Widely used windows are the Bartlett window, the Parzen window, the Tukey-Hanning window, and the Quadratic Spectral (QS) window. For a more detailed account, interested readers can refer to \cite{brockwell2009time} and \cite{priestley1981spectral} in the spectrum estimation literature and \cite{andrews1991heteroskedasticity} in the HAC literature.  

Estimators can also be considered within a parametric family, such as the family of finite-order autoregressive estimators. Define ${X_t}$ as an autoregressive process of order $p$, denoted as $AR(p)$, if there exist constants $\phi_1,...,\phi_p$ such that $$X_t = \phi_1X_{t-1} +...+\phi_pX_{t-p} + \varepsilon_t, \qquad \text{for all} \quad t,$$ 
where $\varepsilon_t \stackrel{i.i.d}{\sim} N (0,\sigma^2)$ and $\varepsilon_t$ is uncorrelated with $X_{t-s}$ for all $s>0$. By estimating the model parameters, $\phi_1,...,\phi_p,\sigma^2$, with $\hat{\phi}_1,...,\hat{\phi}_p,\hat{\sigma}^2$, the spectral density $f(\omega)$ can be estimated using the corresponding autoregressive spectral estimator $$\hat{f}(\omega) = \frac{\hat{\sigma}^2}{2\pi|1-\sum_{k = 1}^p\hat{\phi}_k \exp(i\omega k)|^2}$$ within the family of finite-order autoregressive estimators.
\cite{berk1974consistent} proves that under some weak conditions on $\{X_t\}$ (not assumed to be a finite-order autoregression), if the order of the fitted autoregressive model is asymptotically sufficient to overcome bias, the autoregressive spectrum estimate can yield a consistent estimator of the spectral density of $\{X_t\}$, and he states that the asymptotic variance of the autoregressive spectrum estimator is equivalent to that of the nonparametric
smoothed periodogram estimator. Thus, autoregressive estimates can be used in a nonparametric context. It will be demonstrated later in the simulation section that the parametric autoregressive estimators can be useful in the nonparametric problem of HAC.

When all candidate models are autoregressive models, classical model selection criteria such as Akaike's information criterion (AIC) (See \cite{akaike1974new}), Bayesian information criterion (BIC) (See \cite{schwarz1978estimating}), etc., can be employed. Another consideration is, given the candidate model order $p$, how to estimate the model coefficients. There are many existing methods, including Yule-Walker, Burg, least-squares, maximum likelihood, and restricted maximum likelihood. Those methods will be discussed further in Sections 3 and 4.

As discussed earlier, a truncation parameter must be selected for a nonparametric estimate, and the order of the model needs to be determined for a parametric estimate by the user. To determine which truncation parameter or order is optimal, one might attempt to minimize some criterion that measures the discrepancy between the actual spectral density function and the spectrum estimate. However, in terms of such criteria, the optimal bandwidth of a nonparametric estimate or the optimal order for a parametric estimate depends on the actual spectral density function, which is unknown in practice. Frequency domain cross-validation methods can be used to give a data-driven selection of a spectrum estimate without restricting the form, e.g., parametric or nonparametric. \cite{wahba1975periodic} were the first to use FDCV for selection of tuning constants in a spectral estimator. They focused on spline-based periodogram estimators. The work done by \cite{beltrato1987determining} focuses on nonparametric spectral density estimation based on smoothed periodograms. They show that minimizing mean integrated square error (MISE) is asymptotically equivalent to minimizing a cross-validatory log-likelihood (CVLL). So an optimal bandwidth can be chosen by minimizing CVLL. All candidate estimators in Beltrao and Bloomfield are nonparametric. \cite{robinson1991automatic} considerably extends the results of \cite{beltrato1987determining} by establishing asymptotic properties of the minimizer of CVLL. \cite{hurvich1985data} proposes a unified FDCV method that can select tuning constants for an arbitrary spectral estimator. For example, the method of Hurvich allows for choosing between a parametric estimate and a nonparametric estimate, which it will be shown later has a considerable advantage in the HAC problem. \cite{hurvich1985data} extends the FDCV approach's applicability by introducing a generalized leave-one-out version of the spectral density estimate. Hurvich's purpose is to develop a method that allows researchers to do tuning parameter/model selection across parametric and nonparametric estimates simultaneously. There are two methods introduced in his paper: an autocovariance-based approach and a DFT-based approach. With the autocovariance-based approach, the researcher will be able to do model selection from any spectral density estimate that can be expressed as a function of the sample autocovariances. Those estimates include the lag-weights estimate, discrete periodogram average estimate, and autoregressive estimate using the Yule-Walker method. For example, the autocovariance-based approach allows researchers to choose between a Yule-Walker autoregressive estimate and a periodogram average estimate based on some objective criterion. This approach extended the applicability of the FDCV method developed by Bloomfield and Beltrao, whose approach only allowed for tuning parameter selection within periodogram average estimates. The DFT-based approach of Hurvich is more generally applicable than his autocovariance-based approach. This approach allows the candidates to include any spectral density estimates based on the actual data. Note that not all the spectral density estimates can be expressed in terms of the sample autocovariances, for example an autoregressive estimate based on least-squares or maximum likelihood or restricted maximum likelihood. The DFT-based approach is the one that will be used in the proposed FDCV method for HAC standard error estimation due to its generality. The following is a step-by-step review of this approach.

Hurvich considers the discrete version of $MISE_R$ (Mean Integrated Squared Relative Error) and $MISE_L$ (Mean Integrated Squared Logarithmic Error) as a discrepancy:

$$MISE_R (\hat{f}) = E\Big[\frac{1}{\Tilde{n}}\sum_{j=1}^{\Tilde{n}}\Big(\frac{\hat{f}(\omega_j)-f(\omega_j)}{f(\omega_j)}\Big)^2\Big],$$
$$MISE_L (\hat{f})= E\Big[\frac{1}{\Tilde{n}}\sum_{j=1}^{\Tilde{n}}\Big(\log \hat{f}(\omega_j)-\log f(\omega_j)\Big)^2\Big],$$
where $\Tilde{n}$ is the greatest integer less than or equal to $\frac{n-1}{2}$. For $MISE_R$ and $MISE_L$, the cross-validatory estimates are 
$$CVLL (\hat{f}) = \frac{1}{\Tilde{n}}\sum_{j=1}^{\Tilde{n}}\Big[\log \hat{f}^{-j}(\omega_j)+\frac{I(\omega_j)}{\hat{f}^{-j}(\omega_j)}\Big],$$
$$CV (\hat{f}) =\frac{1}{\Tilde{n}}\sum_{j=1}^{\Tilde{n}}\Big\{\Big[\log \hat{f}^{-j}(\omega_j)-\Big(\log I(\omega_j)+C\Big)\Big]^2-\frac{\pi^2}{6}\Big\},$$
where $C = 0.577216...$ is the Euler constant and $\hat{f}^{-j}$ is the generalized leave-one-out version of the spectral density estimate (see below). Hurvich suggests the use of $CV$ as the selection criterion, and the selected spectral estimator is the one with the lowest $CV$ value. \cite{hurvich1985data} assumes that a zero-mean process generates the data and it is known that the process has zero mean. He defines the leave-one-out version of DFT, $J_k^{-j}$ for $0 \leq k \leq n-1$ and $1 \leq j \leq n$ as
\begin{equation*}
    J_k^{-j} = \begin{cases}
    J_k & \text{if $k \neq j$ and $k \neq n-j$}\\
    \frac{1}{2}(J_{k-1}+J_{k+1}) & \text{if $k = j$ and $k = n-j$}
  \end{cases} 
\end{equation*} 
The inverse Fourier transform can be used to define the leave-one-out version of the data set, ${x_t^{-j}}$, as follows:
$$x_t^{-j} = \sum_{k=0}^{n-1}J_k^{-j}\exp(i\omega_k t), \quad t=0,...,n-1.$$
Based on this leave-one-out version of the data, Hurvich defined the generalized leave-one-out spectral density estimate as
$$\hat{f}^{-j}(\omega_j) = \hat{f}(\omega_j; \{x_t^{-j}\}).$$

\section{A Unified Cross-Validatory Approach to HAC Standard Error Estimation}
\label{sec:meth}
\subsection{A Unified Cross-Validatory Approach to the Estimation of Spectral Density at Zero Frequency}
In Section 2, it was noted that \cite{hurvich1985data} provides the generalized leave-out-one definition of the spectrum estimate $\hat{f}^{-j}(\omega_j) = \hat{f}(\omega_j;\{x_t^{-j}\})$, which can be applied for any spectrum estimate $\hat{f}$. This opens up the possibility for tuning parameter/model selection over a larger class of candidates. This section introduces a cross-validatory approach for HAC standard error estimation by providing a unified truncation parameter/model selection procedure for the spectrum estimate at zero frequency. 

First, the leave-one-out version of the data set must be redefined. The assumption that the mean is known to be zero is a very strong assumption which typically does not hold in practice. Indeed, if this assumption held, then HAC standard errors for $\overline{X_n}$ would not be needed since $\mu$ would be known. Therefore, it is necessary to restrict attention to spectral estimates and cross validation functions that are invariant to the addition of a constant to the data set. Without the zero mean assumption, the DFT-based leave-one-out version of the data is not invariant under adding a constant because $$J_1^{-1} = \frac{1}{2}(J_0 + J_2) \quad \text{and} \quad  J_0 = \frac{1}{n}\sum_{t=0}^{n-1}x_t=\overline{X_n}.$$
Hence, $J_1^{-1}$ is not invariant under adding a constant.
To handle this issue,  the leave-out-one version of the DFT for $j=1,...,n-1$ and $k=1,...,n-1$ is redefined as 
\begin{equation*}
    J_k^{-j} = \begin{cases}
    J_k & \text{if $k \neq j$ and $k \neq n-j$}\\
    \frac{1}{2}(J_{k-1}+J_{k+1}) & \text{if $k = j$ or $k = n-j$ ($j \neq 1$,  $j \neq n-1$}) \\
    J_2 & \text{if $k=1$ and $j=1, n-1$} \\
    J_{n-2} & \text{if $k = n-1$ and $j=1 , n-1$.} \\
    
  \end{cases}
\end{equation*} 
Additionally, it is necessary to adjust the computation of the leave-one-out version of the data set by removing the zero frequency, and therefore, the redefinition is as follows:
$$x_t^{-j} = \sum_{k=1}^{n-1}J_k^{-j}\exp(i\omega_k t).$$
Under this definition, the DFT-based FDCV is invariant under adding a constant.
 
Now, consider an arbitrary class $\textbf{C}$ of candidate spectrum estimates. The estimate from $\textbf{C}$ that minimizes the HAC version of $CV$
$$CV(\hat{f},c) = \frac{1}{\lfloor \tilde{n}^c \rfloor}\sum_{j=1}^{\lfloor \tilde{n}^c \rfloor} \Big\{\Big[log\hat{f}^{-j}(\omega_j)-\Big(logI(\omega_j)+C\Big)\Big]^2-\frac{\pi^2}{6}\Big\}$$ is selected, where $c \in (0,1)$ is a constant. Note that an argument $c$ has been introduced into the $CV$ function. Hurvich's (1985) cross-validation function in this case is $CV(\hat{f},1)$, while it is restricted to $c \in (0,1)$ in this methodology. In practice, it is suggested to take $c =  4/5$. The chosen spectrum estimate will be the unified cross-validatory estimate, and it will be used to estimate the spectral density of the time series at zero frequency.

\subsection{Comparisons with the Method of Hurvich (1985)}
The proposed FDCV method for the HAC problem and the method of \cite{hurvich1985data} differ in the cross-validation function and the candidate class $\textbf{C}$. They also differ in the definition of the leave-one-out version of the data set, as explained in Subsection 3.1.
\subsubsection{Cross-Validation Function}
The cross-validation function $CV(\hat{f},1)$ of \cite{hurvich1985data} is defined over a frequency band that extends from the 1-st to the  $\tilde{n}$-th Fourier frequency and $CV(\hat{f},c)$ is defined over a frequency band between the 1-st to the $(\lfloor\tilde{n}^{c}\rfloor)$-th Fourier frequency for any $c\in (0,1)$. 

Why is such a change essential?  \cite{andrews1991heteroskedasticity} mentioned the potential application of the FDCV method proposed by \cite{beltrato1987determining} to the HAC problem. However, he considered the method to be not well-suited to HAC standard error estimation as the cross-validation criterion $CV(\hat{f},1)$ is a global measure over the frequency band $[0,\pi]$, while the HAC problem focuses on estimating the density at a single frequency, zero. The modification $c \in (0,1)$ makes the criterion function asymptotically local to zero frequency. Indeed, the largest Fourier frequency in $CV(\hat{f},c)$ approaches zero for $c \in (0,1)$,
$$\lim\limits_{n \to \infty}\frac{2\pi (\frac{n}{2})^c}{n} = \lim\limits_{n \to \infty}\pi \left(\frac{n}{2}\right)^{c-1} = 0.$$

Philosophically, the application of FDCV to the HAC problem is motivated by John Tukey's idea of ``borrowing strength". Strength is borrowed from the neighboring frequencies around zero to obtain a stable estimate of $f(0)$.

The constant $c\in (0,1)$ is an arbitrary tuning constant in the proposed method. Based on simulation results, it is suggested to take $c =  4/5$ to handle the bias variance trade-off inherent in the frequency range used in the cross-validation function.  In fact, any values of $c$ between $4/5$ and 1 work quite well based on the simulations not shown here, while choosing a relatively lower value of $c$ will improve the computational speed if the sample size is large. A theoretical justification of $CV(\hat f, c)$ for a given $c \in (0,1)$ is beyond the scope of the current paper. \cite{robinson1991automatic}, who worked with $CVLL$ for lag-weights estimators only, assumed that the cross-validation function is summed over all $n$ Fourier frequencies, corresponding to the global interval $[-\pi,\pi]$. His proof relied in a crucial way on Parseval's formula, which involves summation over all Fourier frequencies, and it is a nontrivial task to derive theoretical properties of local versions of the cross-validation function. Preliminary investigations lead us to the conjecture that for lag-weights spectral estimators, under suitable regularity conditions, the truncation point (or lag number) that minimizes $CV(\hat f, c)$ for a fixed $c \in (1/5,1)$ converges to the truncation point that minimizes the (local) relative mean squared error $E[ (\hat f(0) /f(0) -1)^2]$. This conjecture suggests that one should use a value of $c$ that satisfies $1/5 < c < 1$. The choice used in this paper of $c=4/5$ satisfies this constraint. 

It is worth noting that the problem of choosing $c$ is analogous to the multiplier 2 in the $AIC$ formula ($AIC = 2k - 2\ln(\hat{L})$ where $k$ is the number of estimated parameters in the model and $\hat{L}$ is the maximum value of the likelihood function for the model). However, one can replace $2k$ with $\alpha k$ for $\alpha > 0$. In fact, there does not exist a way to choose that $\alpha$, but any particular fixed $\alpha$ will in some circumstances work well and lead to good asymptotic properties (such as asymptotically efficient model selection). In other words, for any kind of model selection problem or bandwidth selection problem, there is always going to be some tuning constant that one cannot select. Although not directly related to the problem at hand, this analogy helps to understand the role of $c$. To help readers better understand why $c=0.8$ is a reasonable choice, a scenario featuring different values of the exponent $c$ is presented in Subsection 5.6.

\subsubsection{Candidate Class}
The candidates class of \cite{hurvich1985data} includes Daniell average periodogram estimates and Yule-Walker autoregressive estimates. In this paper, the candidate class $\textbf{C}$ is taken to consist of lag-weights estimates with the Parzen kernel and REML autoregressive estimates. 

For the nonparametric candidates, the shift from the Daniell periodogram average estimates to lag-weights estimates with the Parzen kernel is motivated by findings of \cite{newey1987simple},  \cite{newey1994automatic}, \cite{andrews1991heteroskedasticity} and \cite{andrews1992improved}. All those works are based on lag-weights estimators. \cite{newey1994automatic} conclude, based on a simulation, that the effect of the choice of kernel between Bartlett, Parzen, and QS (quadratic spectral) on the performance of the estimator is negligible. The Parzen kernel is used as it never generates negative estimates, unlike Tukey-Hanning. Although the QS kernel was not included in the analysis, it is noteworthy that it yields a non-negative spectral density estimator (See \cite{andrews1991heteroskedasticity}).  For the HAC problem, a negative estimator of $f(0)$ is completely useless for inference as it implies that the estimator of the variance of the sample mean is negative. Compared with the Barlett window, the Parzen estimator has lower asymptotic variance theoretically if the infeasible optimal truncation point were used.

For parametric model candidates, it is proposed to estimate autoregressive models using REML instead of Yule-Walker, MLE or least-squares (least-squares is used to estimate the $AR(1)$ filter in \cite{andrews1992improved} and \cite{newey1994automatic}). The use of REML to estimate the autoregressive model parameters is based on two considerations. First is the reduction of bias of REML compared with other autoregressive estimators. The bias of the Yule-Walker estimator is more prominent than that of other popular autoregressive estimators in small samples. Yule-Walker performs poorly for autoregressive models having a root that is close to the unit circle (or more generally when the spectral density function has strong peaks or troughs). For least-squares and MLE, \cite{cheang2000bias} showed that given an AR(1) process, the bias of estimation of the autoregressive coefficient will be as much as doubled when the root is close to the unit circle. They also showed that the REML estimates of the autoregressive parameters, which do not require knowledge of the mean, perform equivalently up to a term of $O(\frac{1}{n})$ compared to MLE or least-squares with the infeasible knowledge of the mean. Notice that one potential problem of the existing HAC methods is that if the data generating mechanism has a sharp peak in the spectral density at zero frequency, using an $AR(1)$ filter based on the least-squares estimator might fail to adequately prewhiten the data. There is reason to hope that if the autoregressive estimators are based on REML, it will lead to a better prewhitening. This motivates the use of REML to estimate the autoregressive model.

On the other hand, even though the cross-validation procedure  being proposed is invariant under adding a constant (whether or not to use mean-corrected data will not influence the chosen spectral estimator), it is necessary to apply the chosen spectral estimator to mean-corrected data. This is because the lag-weights estimator is not invariant under adding a constant. For parametric estimators, Yule-Walker, least-squares, or MLE estimators are not invariant under adding a constant, so it is necessary to subtract the sample mean when estimating the spectrum with these methods. However, for Yule-Walker, Least-squares, and MLE, demeaning the data will compromise their performance relative to the infeasible case where it is assumed that the population mean is zero, and the method is based on non-mean-adjusted data. One advantage of REML is that it is naturally invariant under adding a constant, which is a desirable property for spectrum estimation of a process with an unknown mean.

REML has been implemented in the context of mixed linear effects models, which can be specialized to yield an autoregressive model. The restricted likelihood function is not well-defined for non-stationary models. A REML autoregressive estimator constrained for stationarity is not implemented in any widely-available packages, as far as it is known. There is also no known discussion of computational efficiency for REML autoregressive estimators. The discussion of how to impose the stationarity constraints as well as the computational implementation of a fast $AR(p)$ REML estimate {will be provided in Section 4. This leads to an algorithm for the evaluation of the autoregressive restricted likelihood that is of interest in its own right. 

\section{The Restricted Likelihood for an Autoregressive Model and Its Efficient Computation}
\label{sec:reml}
\subsection{Restricted  Likelihood  for  an  Autoregressive Model}

Consider an $AR(p)$ process $X_t = \sum_{j=1}^p\phi_jX_{t-j} + \varepsilon_t$ where $\varepsilon_t \stackrel{i.i.d}{\sim} N (0,\sigma^2)$. By \cite{chen2012restricted}, using the \cite{harville1974bayesian} formula, up to an additive constant, based on $X={(X_0,...,X_{n-1})}^T$, the restricted log-likelihood is given by 
\begin{align*}
    L(X,\phi,\sigma^2) &= -\frac{n-1}{2}\log\sigma^2 + \frac{1}{2}\log\frac{|\Sigma_n^{-1}|}{|W^{T}\Sigma_n^{-1}W|}\\
    &-\frac{1}{2\sigma^2}\{X^{T}\Sigma_n^{-1}X-X^{T}\Sigma_n^{-1}W(W^{T}\Sigma_n^{-1}W)^{-1}W^{T}\Sigma_n^{-1}X\},
\end{align*}
where $W = (1,...,1)^T$ and $\Sigma_n = \sigma^{-2}Var(X)$.\\ 
Note that
$$\Sigma_n = \sigma^{-2}\begin{bmatrix}
c_0 & c_1 & c_2 & ... & c_{n-2} & c_{n-1} \\
c_1 & c_0 & c_1 & ... & c_{n-3} & c_{n-2} \\
... & ... & ... & ... & ... & ... \\
c_{n-1} & c_{n-2} & ... & ... & c_1 & c_0 
\end{bmatrix}  $$
is a Toeplitz matrix.

Since the restricted likelihood function is not well-defined for non-stationary models, it is necessary to restrict the search region to the parameter space that will produce a stationary solution. One way to do this is to optimize the restricted likelihood function with respect to the partial correlations (PACF) $\Phi =(\phi_{11},\phi_{22},...,\phi_{pp})$ and perform the optimization of the REML likelihood for $\Phi \in (-1,1)^p$. 
The starting value for REML estimation is taken to be the Burg estimator, as it yields a stationary solution. Then, the Burg autoregressive estimator is transformed to its associated PACF values (For details, see \cite{mcleod2006partial} and their R packages FitAR). The search is performed for $\Phi \in (-1,1)^p$ and $\sigma^2 >0 $, obtaining the $\Phi^{*}$ and $(\sigma^2)^{*}$ that maximize the restricted log likelihood. Afterward, the Durbin-Levinson Recursion is applied to transform the PACF to the autoregressive parameters. This results in the REML estimate that is constrained for stationarity.

\subsection{Computational Efficiency of the Restricted Likelihood}
The terms in the restricted likelihood that would be computationally expensive if computed naively are $|\Sigma_n^{-1}|$, $|W^T\Sigma_n^{-1}W|$ and $U^{T}\Sigma_n^{-1}V$ where $U$ and $V$ are either $(x_1,...,x_n)^T$ or $(1,...,1)^T$. Inverting an $n \times n$ matrix without assuming any structure requires $O(n^3)$ operations. Inversion of an $n \times n$ Toeplitz matrix can be done in $O(n^2)$ operations (See \cite{trench1964algorithm}) using the Durbin-Levinson recursion. According to \cite{barndorff1973parametrization}, $|\sum_n^{-1}| = \Pi_{i=1}^p (1-\phi_{ii}^2)^{i}$. For terms $U^T\Sigma_n^{-1}V$, since naive multiplication of an $n \times n$ matrix with an $n$-dimensional column vector would cost $O(n^2)$ steps, it is proposed to use the PCG (preconditioned conjugate gradient) algorithm to reduce the computational cost to $O(n\log n)$ steps. The PCG algorithm efficently solves the Toeplitz system $Ax = b$, which is equivalent to the computation of $A^{-1}b$ without requiring the computation of $A^{-1}$. The PCG algorithm requires a specification of a circulant preconditioner. The one proposed by T. \cite{chan1988optimal} will be used.

\subsubsection{Speed of Convergence of PCG With T. Chan Preconditioner for $AR(1)$ Covariance Matrix}
Suppose the goal is to solve the linear system $\Sigma_n(f) x = b$ where $\Sigma_n(f)$ is the $n \times n$ Toeplitz covariance matrix of $X=(X_0,\cdots,X_{n-1})^T$ and $\{X_t\}$ is a stationary $AR(p)$ process with spectral density $f$. Note that $\Sigma_n (f)$ is symmetric, and the $j,k$ entry of $\Sigma_n(f)$ is given by
\[
\Sigma_n(f) [j,k] = \int_{-\pi}^{\pi} f(\lambda) \exp\{i(j-k)\lambda\} d\lambda.
\]
Here, $x$ is an $n \times 1$ vector which is the solution to the system (unknown) and $b$ is a known $n \times 1$ vector. The preconditioned conjugate gradient (PCG) algorithm applies the conjugate gradient (CG) algorithm (See \cite{shewchuk1994introduction}) to the preconditioned system $C_n^{-1}(f) \Sigma_n(f) x = C_n^{-1}(f) b$, where $C_n(f)$ is an $n \times n$ circulant matrix with entries that are a function of $f$. The focus is on the PCG algorithm using the circulant preconditioner of T. \cite{chan1988optimal} for obtaining a numerical solution to the original Toeplitz system that is accurate to within a relative tolerance $\epsilon$. The algorithm, which is iterative, is considered to have converged once this relative tolerance has been achieved. Since the T. Chan preconditioner is a circulant, each iteration of the PCG algorithm requires $O(n \log n)$ computations by virtue of the Fast Fourier Transform. The following lemma establishes that the PCG algorithm attains superlinear convergence, that is, the number of iterations required for convergence is $O(1)$ as $n \rightarrow \infty$. Therefore, the total number of computations
required to solve the system using this algorithm is $O(n \log n)$.

\begin{Lem} \label{Superlinear Lemma}
The preconditioned conjugate gradient algorithm based on the T. Chan preconditioner converges superlinearly to the solution of $\Sigma_n(f)x=b$ if $f$ is positive and continuous on $[-\pi,\pi]$.
\end{Lem}

It remains to verify that the spectral density $f$ of a stationary $AR(p)$ process is positive and continuous on $[-\pi,\pi]$, but this follows from the formula
\[
f(\lambda) = \frac{\sigma_{\epsilon}^2/(2\pi)}{|1-\sum_{k=1}^p \alpha_k \exp(-ik\lambda)|^2} \,\,\, .
\]
Indeed, continuity follows from the fact that the denominator is bounded away from zero by virtue of the stationarity assumption assumption that all roots of $1-\sum_{k=1}^p \alpha_k z^k$ are outside the unit circle, and positivity follows from the fact that the denominator is positive and finite since $p<\infty$.

\noindent
\textbf{Proof of Lemma \ref{Superlinear Lemma}:} It is well known that the PCG algorithm using a preconditioner $C_n(f)$ converges superlinearly if the eigenvalues of
$C_n(f)^{-1} \Sigma_n (f)$ are clustered around $1$, that is, if for any $\epsilon >0$, there exist positive integers $N$ and $M$ such that for all $n>N$, at most $M$
eigenvalues of $C_n^{-1}(f) \Sigma_n(f) - I_n$ have absolute value greater than $\epsilon$, where $I_n$ is an $n \times n$ identity matrix.
See pages 1093-1094 of \cite{chan1992circulant} for more discussion on the clustered eigenvalues property. It follows from Theorem 4.4 of \cite{chan2000best} with $r=1$
(corresponding to the T. Chan preconditioner) that if $f$ is positive and continuous on $[-\pi,\pi]$ then the eigenvalues of
$C_n(f)^{-1} \Sigma_n (f)$ are clustered around $1$.  $\Box$

\section{Monte-Carlo Study}
\label{sec:MC}
This section presents the results of a Monte-Carlo study comparing the FDCV and popular existing HAC methods for small sample sizes. The performance of the methods is evaluated based on the coverage rate of the confidence intervals. The first estimator is the cross-validatory estimator described in Section 3. The second estimator is the kernel-based HAC method given by \cite{andrews1992improved} using the quadratic spectral window, and $AR(1)$ specification in their tuning-parameter selection procedure, and the least-squares based $AR(1)$ prewhitening filter. It is called AM-PW, short for Andrews and Monahan's prewhitening based HAC estimator. The third estimator is the kernel-based HAC estimator given by \cite{newey1994automatic} using the Bartlett window as described in \cite{newey1987simple}, \cite{newey1994automatic}, a nonparametric truncation parameter selection procedure, and the least-squares-based $AR(1)$ prewhitening filter. It is referred to NW-PW, short for Newey and West prewhitening based estimator. For a more detailed step-by-step procedure, readers can refer to \cite{den199712}. Several experiments suggested by \cite{andrews1992improved} and \cite{den199712} are investigated, considering two sample sizes, $n = 50, 200$, and reporting the coverage rates of the HAC-based confidence intervals at nominal rates $90\%$, $95\%$ and $99\%$ based on 3000 replications. The computations for the AM-PW and NW-PW methods are based on the Sandwich package in R (See \cite{sandwich}). 

In the FDCV method for HAC, the class $\textbf{C}$ of candidate spectrum estimates consisted of the REML-based autoregressive estimates of order 0 to 5 and lag-weights estimates with Parzen kernels from truncation point 1 to $\lfloor4(\frac{n}{100})^\frac{2}{9}\rfloor$. For each realization, and for each of the candidates $\hat{f}$ in $\textbf{C}$, The respective $CV(\hat{f},c)$ is calculated with $c=4/5$. 

Define $CV_{AR}(\hat{f}) = CV(\hat{f},4/5)$ when $\hat{f}$ is a REML-based autoregressive estimate, and \newline
$CV_{PZ}(\hat{f}) = CV(\hat{f},4/5)$ when $\hat{f}$ is a lag-weights estimate with the Parzen window. The notation $CV_C(\hat{f}) = CV(\hat{f},4/5)$ is reserved for cases when $f$ can be chosen from the combined set of REML-based autoregressive estimates and lag-weights estimates with the Parzen window.

For $CV_{AR}$, $CV_{PZ}$ and $CV_{C}$, after obtaining the selected estimator, the data set $\{x_t\}{_{t=0}^{n-1}}$ is demeaned and the selected estimator is applied to} the mean-corrected time series for estimation at zero frequency. This approach results in obtaining the respective spectral density estimator at zero frequency. For any selected estimator $\hat{f}$, define the estimated standard error for $\hat{\mu} = \overline{X}_n$ as 
$$\hat{\sigma}(\hat{f}) = \sqrt{\frac{2\pi \hat{f}^(0)}{n-1}},$$ 
where $n-1$ in the denominator is a finite-sample correction (See \cite{andrews1991heteroskedasticity}). Using the standard error, the nominal $90\%$, $95\%$ and $99\%$ confidence intervals for $\mu$ are computed, and the observed coverage rates for $CV_{AR}$, $CV_{PZ}$, and $CV_C$ are reported. Notice that for AM-PW and NW-PW, the results will be directly used from the Sandwich package, which provides the standard error directly.

After the observed coverage rates for each method are obtained, the relative efficiency measures for $CV_C$, AM-PW, and NW-PW are provided for data-generating mechanisms at the nominal $95\%$ coverage rate where some methods under-cover while others over-cover. This is intended to give readers an evaluation metric that distinguishes between under-coverage and over-coverage. The relative efficiency is a number in $[0,1]$. In each case, the method with the relative efficiency of $1$ is the one with the best performance, and the method with the lowest relative efficiency is the one with the worst performance. To compute the relative efficiency, it is necessary to construct a measure for badness. The badness $B$ of the actual coverage rate $p$ for the nominal coverage rate $95\%$ is defined as
$$ B(p) = \left\{
\begin{array}{rcl}
 2|logit(p)-logit(0.95)| & &\text{if}\quad p \leq 0.95 \\
|logit(p)-logit(0.95)|& &\text{if} \quad p > 0.95
\end{array} \right. $$
where $logit(p) = \log(\frac{p}{1-p})$. 

In practice, under-coverage is considered to be worse than over-coverage. To take this asymmetry into account, a factor of 2 is used to penalize under-coverage in $B(p)$, when $p \leq 0.95$. 

Denoting $p_1$, $p_2$ and $p_3$ to be the actual coverage probabilities of $CV_C$, AM-PW, NW-PW respectively, then the relative efficiency of $p_i$ can be calculated as $$e(p_i) = \frac{min\{B(p_1),B(p_2),B(p_3)\}}{B(p_i)},\qquad i\in\{1,2,3\}.$$
Note that $e(p_i) \in [0,1]$. If $p_i = 1$, then $logit(p_i) = \infty$. In this case, a value of 0 will be assigned to $e(p_i)$.

Five sets of experiments were conducted. The data-generating mechanisms for Subsection 5.1 are $AR(1)$ processes. For Subsection 5.2, the data-generating mechanism is a white noise process. For Subsection 5.3, they are $MA(1)$ processes. The data-generating mechanisms in Subsections 5.4 and 5.5 were proposed by \cite{den199712}. It is important to note that the primary comparison is between the results of $CV_C$, AM-PW, and NW-PW. The criteria $CV_{AR}$ and $CV_{PZ}$ are used for supplementary analysis. 
In Subsection 5.6, there is a follow-up discussion to Section 3.2.1, with a focus on the role of exponent $c$ in the criterion function. This section offers further clarity and context regarding its importance in the paper.
\newpage
\subsection{$AR(1)$ Processes}
$$X_t = \phi_1X_{t-1}+\varepsilon_t,\qquad \varepsilon_t \stackrel{i.i.d}{\sim} N (0,1)$$
\begingroup
\renewcommand\arraystretch{0.5}
{\begin{longtable}[c]{@{}clcccccc@{}}
\multicolumn{1}{l}{}  &        & \multicolumn{3}{c}{$n = 50$}              & \multicolumn{3}{c}{$n = 200$} \\* \midrule
\endfirsthead
\endhead
\bottomrule
\endfoot
\endlastfoot
\textbf{$\phi_1$} & \textbf{Method} & \textbf{90\%} & \textbf{95\%} & \multicolumn{1}{c}{\textbf{99\%}} & \textbf{90\%} & \textbf{95\%} & \textbf{99\%} \\* \midrule
\multirow{5}{*}{0.1}  & $CV_C$    & 86.7 & 91.9 & \multicolumn{1}{c}{97.4} & 87.3    & 93.1    & 98.3    \\
                      & $CV_{AR}$ & 86.7 & 91.9 & \multicolumn{1}{c}{97.4} & 87.8    & 93.3    & 98.3    \\
                      & $CV_{PZ}$   & 87.2 & 92.5 & \multicolumn{1}{c}{97.8} & 86.9    & 92.7    & 98.4    \\
                      & AM-PW  & 87.7 & 93.1 & \multicolumn{1}{c}{97.7} & 89.6    & 94.8    & 99.0    \\
                      & NW-PW  & 85.3 & 90.8 & \multicolumn{1}{c}{96.4} & 88.4    & 94.2    & 98.8    \\* \midrule
\multirow{5}{*}{0.3}  & $CV_C$    & 81.5 & 88.7 & \multicolumn{1}{c}{94.9} & 85.2    & 91.3    & 97.2    \\
                      & $CV_{AR}$ & 82.1 & 88.6 & \multicolumn{1}{c}{94.6} & 87.3    & 92.8    & 97.8    \\
                      & $CV_{PZ}$   & 80.6 & 87.9 & \multicolumn{1}{c}{95.2} & 81.2    & 88.0    & 95.7    \\
                      & AM-PW  & 86.5 & 92.0 & \multicolumn{1}{c}{97.2} & 89.6    & 94.6    & 98.9    \\
                      & NW-PW  & 85.2 & 90.6 & \multicolumn{1}{c}{96.3} & 88.6    & 94.1    & 98.8    \\* \midrule
\multirow{5}{*}{0.5}  & $CV_C$    & 79.1 & 85.3 & \multicolumn{1}{c}{92.8} & 84.7    & 91.0    & 96.7    \\
                      & $CV_{AR}$ & 80.3 & 86.2 & \multicolumn{1}{c}{92.8} & 87.9    & 93.1    & 97.7    \\
                      & $CV_{PZ}$   & 76.3 & 82.9 & \multicolumn{1}{c}{91.6} & 77.3    & 84.4    & 92.9    \\
                      & AM-PW  & 85.4 & 90.6 & \multicolumn{1}{c}{96.3} & 89.0    & 94.1    & 98.8    \\
                      & NW-PW  & 84.4 & 89.9 & \multicolumn{1}{c}{95.7} & 88.5    & 93.8    & 98.7    \\* \midrule
\multirow{5}{*}{0.7}  & $CV_C$    & 75.1 & 81.8 & \multicolumn{1}{c}{89.9} & 85.0    & 90.5    & 96.6    \\
                      & $CV_{AR}$ & 79.2 & 84.5 & \multicolumn{1}{c}{90.9} & 87.5    & 92.6    & 97.6    \\
                      & $CV_{PZ}$   & 68.7 & 76.7 & \multicolumn{1}{c}{87.3} & 79.4    & 85.8    & 93.2    \\
                      & AM-PW  & 82.7 & 88.1 & \multicolumn{1}{c}{94.3} & 88.0    & 93.4    & 98.3    \\
                      & NW-PW  & 81.8 & 87.5 & \multicolumn{1}{c}{94.0} & 87.9    & 92.9    & 98.4    \\* \midrule
\multirow{5}{*}{0.9}  & $CV_C$    & 70.8 & 77.2 & \multicolumn{1}{c}{84.7} & 84.1    & 89.4    & 94.8    \\
                      & $CV_{AR}$ & 75.7 & 81.5 & \multicolumn{1}{c}{88.0} &85.6    & 90.8    & 95.7    \\
                      & $CV_{PZ}$   & 46.6 & 53.6 & \multicolumn{1}{c}{66.7} & 67.2    & 74.8    & 86.2    \\
                      & AM-PW  & 71.7 & 77.6 & \multicolumn{1}{c}{86.9} & 84.8    & 89.6    & 95.9    \\
                      & NW-PW  & 71.1 & 76.9 & \multicolumn{1}{c}{86.5} & 84.7    & 89.4    & 95.9    \\* \midrule
\multirow{5}{*}{0.95} & $CV_C$    & 68.6 & 74.4 & \multicolumn{1}{c}{82.4} & 82.0   & 87.3    & 93.4    \\
                      & $CV_{AR}$ & 73.2 & 78.8 & \multicolumn{1}{c}{85.7} & 82.8    & 88.1    & 94.1    \\
                      & $CV_{PZ}$   & 33.4 & 39.8 & \multicolumn{1}{c}{51.2} & 53.1    & 60.4    & 73.0    \\
                      & AM-PW  & 62.7 & 70.3 & \multicolumn{1}{c}{79.5} & 79.9    & 86.2    & 92.9    \\
                      & NW-PW  & 62.2 & 69.9 & \multicolumn{1}{c}{79.0} & 79.9    & 86.0    & 92.8    \\* \bottomrule
\caption{Coverage Rates for $AR(1)$ Processes}
\label{table1}\\
\end{longtable}}
\endgroup

\newpage

\vskip -0.8cm
For the $AR(1)$ process, the peak at zero frequency becomes sharper when the $AR(1)$ coefficient $\phi_1$ increases. In all the cases, the actual coverage probabilities are smaller than the nominal coverage probabilities. When $\phi=0.1, 0.3, 0.5, 0.7, 0.9$, the method that under-covers the least is AM-PW for both $n=50$ and $200$. For $\phi_1 = 0.95$, the method that under-covers the least is $CV_C$ for both $n=50$ and $n=200$. 

For $\phi=0.1, 0.3, 0.5, 0.7, 0.9$, the prewhitening-based method is superior because the order of the autoregressive prewhitening filter is the same as the order of the autoregressive data-generating process. In other words, this is precisely the process that AM-PW and NW-PW are built for. It is observed that even with $CV_{AR}$, where the choice of candidates is restricted to autoregressive models, $CV_{AR}$ is still outperformed by AM-PW and NW-PW. One explanation for this is that $CV_{AR}$ does not always choose the true order (In this case, 1) especially in small samples. However, as will be seen in subsection 5.5, the lack of model selection in the prewhitening filter can lead to undesirable performance in terms of coverage rates. 

The bandwidth selection procedure of AM-PW follows the proposal of  \cite{andrews1991heteroskedasticity} and works well when the prewhitened process has a monotonically decreasing spectral density. The least-squares $AR(1)$ estimator tends to yield an $AR(1)$ coefficient that is biased downward. Therefore, the spectral density of the prewhitened time series based on the least-squares $AR(1)$ filter can still be monotonically decreasing, which favors AM-PW.

Even though AM-PW and NW-PW are built for $AR(1)$ processes due to $AR(1)$ filter application in both methods, they are still outperformed by $CV_C$ in the case when $\phi = 0.95$ and by $CV_{AR}$ when $\phi = 0.9$ and 0.95. These simulation results support the motivation behind the application of REML to estimate autoregressive processes in the HAC problem. Hence, it is suggested to use REML whenever estimating an autoregressive model. 

It is worth mentioning that if the data generating process has a peaked spectral density at zero frequency, an autoregressive estimator is chosen more often by $CV_C$. For example, when $n=50$ and $\phi_1 = 0.9$, an autoregressive estimator is chosen by $CV_C$ $75.2\%$ of the time, and $CV_{AR}$ is able to pick out the true lag length for AR(1) models $69.2\%$ of the time. The choice becomes far more accurate with a larger sample size. Interestingly, this observation is not limited to autoregressive data-generating processes. In fact, it applies to any data-generating process that yields a peaked spectral density at zero frequency. For example, an experiment with a sample size of 50 is conducted using an ARMA(1,1) process, defined as $X_t = 0.9X_{t-1} -0.4\varepsilon_{t-1} + \varepsilon_t$, where $\varepsilon_t \stackrel{i.i.d}{\sim} N(0,1)$. Over 3000 replications, it is observed that the autoregressive estimator was chosen by $CV_C$ $69.0\%$ of the time. However, if the spectral density appears to be flatter, a lag-weights estimate with the Parzen window will be chosen more often compared to the DGP where the true spectral density has a peak at zero frequency. When the DGP is a causal and invertible ARMA(1,1) process $X_t = 0.2X_{t-1} -0.2\varepsilon_{t-1} + \varepsilon_t$ with $\varepsilon_t \stackrel{i.i.d}{\sim} N(0,1)$ and a sample size of 50 over 3000 replications, it is observed that lag-weights estimate with the Parzen window is chosen by $CV_C$ $54.0\%$ of times.

Finally, the gap between the observed coverage rate and the nominal coverage rate of $CV_C$ has narrowed when $n$ moves from 50 to 200.

\subsection{White Noise Process}

$$X_t = \varepsilon_t,\qquad \varepsilon_t \stackrel{i.i.d}{\sim} N (0,1)$$
\begingroup
\renewcommand\arraystretch{0.5}
\begin{longtable}[c]{@{}clccccllll@{}}
\toprule
\textbf{$n$} &
  \textbf{Method} &
  \textbf{90\%} &
  \textbf{95\%} &
  \textbf{99\%} &
  \textbf{n} &
  \textbf{Method} &
  \textbf{90\%} &
  \textbf{95\%} &
  \textbf{99\%} \\* \midrule
\endfirsthead
\endhead
\bottomrule
\endfoot
\endlastfoot
\multirow{5}{*}{50} &
  $CV_C$ &
  88.8 &
  93.4 &
  98.3 &
  \multirow{5}{*}{200} &
  $CV_C$ &
  89.5 &
  94.6 &
  99.0 \\
 & $CV_{AR}$ & 88.9 & 93.5 & 98.3 &  & $CV_{AR}$ & 89.7 & 94.6 & 99.1 \\
 & $CV_{PZ}$   & 89.7 & 94.5 & 98.8 &  & $CV_{PZ}$   & 89.9 & 94.9 & 99.2 \\
 & AM-PW  & 88.1 & 93.1 & 97.9 &  & AM-PW  & 89.7 & 94.7 & 99.0 \\
 & NW-PW  & 85.5 & 90.9 & 96.4 &  & NW-PW  & 88.4 & 94.1 & 98.8 \\* \bottomrule
\caption{Coverage Rates for White Noise Processes}
\label{table3}\\
\end{longtable}
\endgroup

For the white noise process, the spectral density function is flat. For both $n=50$ and $200$, the actual coverage probabilities are smaller than the nominal coverage probabilities. The methods that perform the best between $CV_C$, AM-PW, and NW-PW for $95\%$ nominal coverage rate are $CV_C$ for $n=50$ and AM-PW with a slight advantage when $n=200$ (though based on Table 2, all methods provide satisfactory performance when $n=200$). For $CV_C$, these results show the advantage of the inclusion of nonparametric spectrum estimates, as seen from the table that $CV_{PZ}$ has better performance than $CV_{AR}$. The only method that has difficulties in this experiment is NW-PW. Its under-coverage of the confidence interval is substantial when $n=50$. Finally, the gap between the observed coverage rate and the nominal coverage rate of $CV_C$ has narrowed when $n$ moves from 50 to 200.

\subsection{$MA(1)$ Processes }
$$X_t = \varepsilon_t + \psi_1\varepsilon_{t-1},\qquad \varepsilon_t \stackrel{i.i.d}{\sim} N (0,1).$$

\begingroup
\renewcommand\arraystretch{0.5}
\begin{longtable}[c]{@{}clcccccc@{}}
\textbf{}       & \textbf{}          & \multicolumn{3}{c}{\textbf{$n = 50$}}           & \multicolumn{3}{c}{\textbf{$n = 200$}}         \\* \midrule
\endfirsthead
\endhead
\bottomrule
\endfoot
\endlastfoot
\textbf{$\psi_1$} & \textbf{Method} & \textbf{90\%} & \textbf{95\%} & \textbf{99\%} & \textbf{90\%} & \textbf{95\%} & \textbf{99\%} \\* \midrule
\multirow{5}{*}{-0.3} & $CV_C$    & 92.0 & 95.3 & 98.4  & 90.2  & 95.2  & 99.1  \\
                      & $CV_{AR}$ & 92.4 & 95.5 & 98.3  & 90.2  & 95.2  & 99.0  \\
                      & $CV_{PZ}$   & 95.2 & 97.7 & 99.5  & 96.0  & 98.3  & 99.8  \\
                      & AM-PW  & 92.2 & 95.9 & 99.1  & 93.1  & 97.0  & 99.7  \\
                      & NW-PW  & 86.0 & 90.9 & 96.4  & 89.4  & 94.5  & 98.9  \\* \midrule
\multirow{5}{*}{-0.5} & $CV_C$    & 92.0 & 95.6 & 98.7  & 91.7  & 95.8  & 99.1  \\
                      & $CV_{AR}$ & 91.9 & 95.4 & 98.6  & 91.7  & 95.8  & 99.1  \\
                      & $CV_{PZ}$   & 97.2 & 98.7 & 99.7  & 97.0  & 98.5  & 99.8  \\
                      & AM-PW  & 96.6 & 98.5 & 99.7  & 97.7  & 99.4  & 100.0 \\
                      & NW-PW  & 84.7 & 89.7 & 95.3  & 89.4  & 94.5  & 98.6  \\* \midrule
\multirow{5}{*}{-0.7} & $CV_C$    & 95.8 & 97.7 & 99.3  & 94.8  & 97.9  & 99.7  \\
                      & $CV_{AR}$ & 95.7 & 97.7 & 99.3  & 95.0  & 98.0  & 99.7  \\
                      & $CV_{PZ}$   & 99.2 & 99.8 & 100.0 & 98.4  & 99.4  & 100.0 \\
                      & AM-PW  & 99.5 & 99.9 & 100.0 & 100.0 & 100.0 & 100.0 \\
                      & NW-PW  & 85.5 & 91.0 & 95.4  & 87.5  & 92.4  & 96.6  \\* \bottomrule
\caption{Coverage Rates for $MA(1)$ Processes}
\label{table5}\\
\end{longtable}
\endgroup

\begingroup
\renewcommand\arraystretch{0.5}
\begin{longtable}[c]{@{}clcc@{}}
\toprule
\textbf{$\psi_1$}                   & \textbf{Method} & \textbf{$n = 50$} & \textbf{$n = 200$} \\* \midrule
\endfirsthead
\endhead
\bottomrule
\endfoot
\endlastfoot
\multirow{3}{*}{-0.3}             & $CV_C$                & 1.00            & 1.00             \\
                                  & AM-PW              & 0.31            & 0.08             \\
                                  & NW-PW              & 0.05            & 0.23             \\* \midrule
\multirow{3}{*}{-0.5}             & $CV_C$                & 1.00            & 1.00             \\
                                  & AM-PW              & 0.11            & 0.09             \\
                                  & NW-PW              & 0.09            & 1.00             \\* \midrule
\multirow{3}{*}{-0.7}             & $CV_C$                & 1.00            & 1.00             \\
                                  & AM-PW              & 0.20            & 0.00             \\
                                  & NW-PW              & 0.64            & 0.99             \\* \bottomrule
\caption{$MA(1)$ Process Relative Efficiency}
\label{table6}\\  
\end{longtable}
\endgroup

For all values of $\psi_1$, the intervals based on $CV_C$ and AM-PW overcover $\mu$ and those based on NW-PW undercover $\mu$. Thus, referring to relative efficiency for evaluation is necessary. When $\psi_1 = -0.3, -0.5,-0.7$, $CV_C$ has more reliable performance than AM-PW and NW-PW.

Different values of the $MA(1)$ coefficient are taken. When the $MA(1)$ coefficient $\psi_1$ approaches -1, $f(0)$ approaches 0. This provides an alternative way of doing stress-testing for the proposed method and traditional HAC methods, as there will be a trough in the spectral density at zero frequency when $\psi_1$ is close to -1. Note that when $f(0)= 0$, the assumption of short memory will no longer hold true, and the HAC standard error will not be consistent. When $\psi_1 = -0.7$, AM-PW have a coverage probability of $100\%$ even for a $90\%$ confidence interval when $n=200$.

Finally, the gap between the observed and nominal coverage rates for $CV_C$ is narrowed when $n$ moves from 50 to 200 in all cases expect when $\psi_1$ is close to -1.

\subsection{$MA(2)$ and $MA(3)$ Processes}
$$X_t = \varepsilon_t + \alpha\varepsilon_{t-1}+\beta\varepsilon_{t-q},\qquad q\in\{2,3\} \quad \text{and} \quad \varepsilon_{t-q} \stackrel{i.i.d}{\sim} N (0,1)$$

\begingroup
\renewcommand\arraystretch{0.5}
\begin{longtable}[c]{@{}llllcccllccc@{}}
\toprule
\textbf{$\alpha$} &
  \textbf{$\beta$} &
  \textbf{$q$} &
  \textbf{Method} &
  \textbf{90\%} &
  \textbf{95\%} &
  \textbf{99\%} &
  \textbf{$q$} &
  \textbf{Method} &
  \textbf{90\%} &
  \textbf{95\%} &
  \textbf{99\%} \\* \midrule
\endfirsthead
\multicolumn{12}{c}%
{{\bfseries Table \thetable\ continued from previous page}} \\
\endhead
\bottomrule
\endfoot
\endlastfoot
\multirow{5}{*}{0.0}  & \multirow{5}{*}{-0.3} & \multirow{5}{*}{2} & $CV_C$    & 93.2 & 96.6 & 99.0 & \multirow{5}{*}{3} & $CV_C$    & 95.7 & 97.7 & 99.4 \\
                      &                       &                    & $CV_{AR}$ & 93.2 & 96.5 & 99.0 &                    & $CV_{AR}$ & 95.8 & 97.7 & 99.4 \\
                      &                       &                    & $CV_{PZ}$   & 96.7 & 98.7 & 99.8 &                    & $CV_{PZ}$   & 97.6 & 99.2 & 99.9 \\
                      &                       &                    & AM-PW  & 97.2 & 99.0 & 99.8 &                    & AM-PW  & 96.9 & 98.8 & 99.8 \\
                      &                       &                    & NW-PW  & 86.1 & 90.3 & 95.5 &                    & NW-PW  & 95.3 & 97.8 & 99.6 \\* \midrule
\multirow{5}{*}{-0.1} & \multirow{5}{*}{-0.3} & \multirow{5}{*}{2} & $CV_C$    & 94.0 & 96.9 & 99.2 & \multirow{5}{*}{3} & $CV_C$    & 96.8 & 98.2 & 99.4 \\
                      &                       &                    & $CV_{AR}$ & 94.0 & 96.7 & 99.2 &                    & $CV_{AR}$ & 96.5 & 98.1 & 99.4 \\
                      &                       &                    & $CV_{PZ}$   & 97.8 & 99.2 & 99.9 &                    & $CV_{PZ}$   & 99.0 & 99.7 & 99.9 \\
                      &                       &                    & AM-PW  & 98.2 & 99.4 & 99.9 &                    & AM-PW  & 98.7 & 99.6 & 99.9 \\
                      &                       &                    & NW-PW  & 86.0 & 90.5 & 95.9 &                    & NW-PW  & 96.4 & 98.4 & 99.7 \\* \midrule
\multirow{5}{*}{0.0}  & \multirow{5}{*}{0.3}  & \multirow{5}{*}{2} & $CV_C$    & 83.3 & 89.4 & 95.6 & \multirow{5}{*}{3} & $CV_C$    & 82.3 & 89.1 & 95.3 \\
                      &                       &                    & $CV_{AR}$ & 83.7 & 89.4 & 95.9 &                    & $CV_{AR}$ & 82.8 & 89.6 & 95.6 \\
                      &                       &                    & $CV_{PZ}$   & 83.4 & 89.7 & 96.1 &                    & $CV_{PZ}$   & 81.5 & 88.6 & 95.6 \\
                      &                       &                    & AM-PW  & 78.5 & 85.6 & 93.7 &                    & AM-PW  & 78.8 & 86.0 & 93.8 \\
                      &                       &                    & NW-PW  & 81.7 & 87.6 & 95.0 &                    & NW-PW  & 77.4 & 84.4 & 92.3 \\* \midrule
\multirow{5}{*}{0.1}  & \multirow{5}{*}{0.3}  & \multirow{5}{*}{2} & $CV_C$    & 82.2 & 88.5 & 95.3 & \multirow{5}{*}{3} & $CV_C$    & 81.2 & 87.6 & 94.2 \\
                      &                       &                    & $CV_{AR}$ & 82.0 & 88.5 & 95.0 &                    & $CV_{AR}$ & 81.5 & 87.7 & 94.4 \\
                      &                       &                    & $CV_{PZ}$   & 82.0 & 88.7 & 95.4 &                    & $CV_{PZ}$   & 80.2 & 87.2 & 94.3 \\
                      &                       &                    & AM-PW  & 80.0 & 86.7 & 94.3 &                    & AM-PW  & 76.8 & 84.2 & 92.4 \\
                      &                       &                    & NW-PW  & 82.8 & 88.3 & 95.2 &                    & NW-PW  & 76.2 & 83.4 & 92.0 \\* \bottomrule
\caption{Coverage Rates for $MA(2)$ and $MA(3)$ Processes: $n = 50$}
\label{table7}\\
\end{longtable}
\endgroup

\begingroup
\renewcommand\arraystretch{0.5}
\begin{longtable}[c]{@{}llllcccllccc@{}}
\toprule
\textbf{$\alpha$} &
  \textbf{$\beta$} &
  \textbf{$q$} &
  \textbf{Method} &
  \textbf{90\%} &
  \textbf{95\%} &
  \textbf{99\%} &
  \textbf{$q$} &
  \textbf{Method} &
  \textbf{90\%} &
  \textbf{95\%} &
  \textbf{99\%} \\* \midrule
\endfirsthead
\multicolumn{12}{c}%
{{\bfseries Table \thetable\ continued from previous page}} \\
\endhead
\bottomrule
\endfoot
\endlastfoot
\multirow{5}{*}{0.0}  & \multirow{5}{*}{-0.3} & \multirow{5}{*}{2} & $CV_C$    & 91.7 & 95.9  & 99.0  & \multirow{5}{*}{3} & $CV_C$    & 93.4 & 97.0  & 99.4  \\
                      &                       &                    & $CV_{AR}$ & 91.7 & 95.8  & 99.0  &                    & $CV_{AR}$ & 93.3 & 97.0  & 99.4  \\
                      &                       &                    & $CV_{PZ}$   & 96.9 & 98.8  & 99.8  &                    & $CV_{PZ}$   & 97.5 & 99.0  & 99.9  \\
                      &                       &                    & AM-PW  & 98.6 & 99.7  & 100.0 &                    & AM-PW  & 98.6 & 99.8  & 100.0 \\
                      &                       &                    & NW-PW  & 89.7 & 94.3  & 98.6  &                    & NW-PW  & 89.8 & 94.1  & 98.5  \\* \midrule
\multirow{5}{*}{-0.1} & \multirow{5}{*}{-0.3} & \multirow{5}{*}{2} & $CV_C$    & 92.4 & 96.0  & 99.1  & \multirow{5}{*}{3} & $CV_C$    & 94.3 & 97.5  & 99.5  \\
                      &                       &                    & $CV_{AR}$ & 92.2 & 96.1  & 99.1  &                    & $CV_{AR}$ & 94.3 & 97.5  & 99.5  \\
                      &                       &                    & $CV_{PZ}$   & 97.7 & 98.9  & 99.9  &                    & $CV_{PZ}$   & 98.3 & 99.3  & 100.0 \\
                      &                       &                    & AM-PW  & 99.4 & 100.0 & 100.0 &                    & AM-PW  & 99.8 & 100.0 & 100.0 \\
                      &                       &                    & NW-PW  & 89.7 & 94.0  & 98.4  &                    & NW-PW  & 88.1 & 92.9  & 97.4  \\* \midrule
\multirow{5}{*}{0.0}  & \multirow{5}{*}{0.3}  & \multirow{5}{*}{2} & $CV_C$    & 86.4 & 91.9  & 97.6  & \multirow{5}{*}{3} & $CV_C$    & 87.8 & 93.1  & 98.0  \\
                      &                       &                    & $CV_{AR}$ & 86.7 & 92.0  & 97.5  &                    & $CV_{AR}$ & 88.6 & 93.4  & 98.1  \\
                      &                       &                    & $CV_{PZ}$   & 84.2 & 90.1  & 97.1  &                    & $CV_{PZ}$   & 85.1 & 91.2  & 97.6  \\
                      &                       &                    & AM-PW  & 80.7 & 87.8  & 95.7  &                    & AM-PW  & 81.0 & 87.9  & 95.9  \\
                      &                       &                    & NW-PW  & 86.1 & 91.8  & 97.9  &                    & NW-PW  & 85.2 & 91.1  & 97.6  \\* \midrule
\multirow{5}{*}{0.1}  & \multirow{5}{*}{0.3}  & \multirow{5}{*}{2} & $CV_C$    & 85.7 & 91.4  & 97.3  & \multirow{5}{*}{3} & $CV_C$    & 87.9 & 93.4  & 97.9  \\
                      &                       &                    & $CV_{AR}$ & 86.4 & 91.6  & 97.4  &                    & $CV_{AR}$ & 88.8 & 93.7  & 98.1  \\
                      &                       &                    & $CV_{PZ}$   & 82.0 & 88.4  & 96.1  &                    & $CV_{PZ}$   & 84.4 & 90.7  & 97.0  \\
                      &                       &                    & AM-PW  & 81.9 & 88.8  & 96.3  &                    & AM-PW  & 79.3 & 86.0  & 94.7  \\
                      &                       &                    & NW-PW  & 86.3 & 92.1  & 98.1  &                    & NW-PW  & 85.0 & 91.1  & 97.3  \\* \bottomrule
\caption{Coverage Rates for $MA(2)$ and $MA(3)$ Processes: $n = 200$}
\label{table8}\\
\end{longtable}
\endgroup

\newpage 

\begingroup
\renewcommand\arraystretch{0.5}
\begin{longtable}[c]{@{}ccclccccc@{}}
\toprule
\textbf{$\alpha$} &
  \multicolumn{1}{c}{\textbf{$\beta$}} &
  \textbf{$q$} &
  \textbf{Method} &
  \textbf{$n = 50$} &
  \multicolumn{1}{c}{\textbf{$n = 200$}} &
  \textbf{$q$} &
  \textbf{$n = 50$} &
  \textbf{$n = 200$} \\* \midrule
\endfirsthead
\endhead
\bottomrule
\endfoot
\endlastfoot
0.0        & \multicolumn{1}{c}{-0.3}       & 2      & $CV_C$   & 1.00 & \multicolumn{1}{c}{1.00} & 3 & 1.00 & 0.64 \\
           & \multicolumn{1}{c}{}           &        & AM-PW & 0.26 & \multicolumn{1}{c}{0.07} &   & 0.56 & 0.10 \\
           & \multicolumn{1}{c}{}           &        & NW-PW & 0.29 & \multicolumn{1}{c}{0.74} &   & 0.95 & 1.00 \\* \midrule
-0.1       & \multicolumn{1}{c}{-0.3}       & 2      & $CV_C$   & 1.00 & \multicolumn{1}{c}{1.00} & 3 & 1.00 & 1.00 \\
           & \multicolumn{1}{c}{}           &        & AM-PW & 0.23 & \multicolumn{1}{c}{0.05} &   & 0.42 & 0.15 \\
           & \multicolumn{1}{c}{}           &        & NW-PW & 0.37 & \multicolumn{1}{c}{0.58} &   & 0.91 & 0.98 \\* \midrule
0.0        & \multicolumn{1}{c}{0.3}        & 2      & $CV_C$   & 1.00 & \multicolumn{1}{c}{1.00} & 3 & 1.00 & 1.00 \\
           & \multicolumn{1}{c}{}           &        & AM-PW & 0.70 & \multicolumn{1}{c}{0.52} &   & 0.75 & 0.35 \\
           & \multicolumn{1}{c}{}           &        & NW-PW & 0.82 & \multicolumn{1}{c}{0.97} &   & 0.67 & 0.56 \\* \midrule
0.1        & \multicolumn{1}{c}{0.3}        & 2      & $CV_C$   & 1.00 & \multicolumn{1}{c}{0.85} & 3 & 1.00 & 1.00 \\
           & \multicolumn{1}{c}{}           &        & AM-PW & 0.84 & \multicolumn{1}{c}{0.57} &   & 0.77 & 0.26 \\
           & \multicolumn{1}{c}{}           &        & NW-PW & 0.97 & \multicolumn{1}{c}{1.00} &   & 0.74 & 0.47 \\* \bottomrule

\caption{$MA(2)$ and $MA(3)$ Process Relative Efficiency}
\label{table9}\\
\end{longtable}

\endgroup

\cite{den199712} suggest these generating mechanisms to compare different HAC estimators' robustness against various autocorrelation structures. The parameters are chosen such that the first-order autocorrelation for the prewhitened time series is small, but the higher-order autocorrelations are substantial. Compared with AM-PW and NW-PW, $CV_C$ is superior in most situations and yields the best overall performance in this experiment. If the MA coefficient $\beta$ is negative, then AM-PW tends to lead to substantial over-coverage of $\mu$, but NW-PW tends to under-cover $\mu$. If the MA coefficient $\beta$ is positive, then both AM-PW and NW-PW tend to substantially undercover $\mu$. $CV_C$ generally has better performance when $\beta$ is negative and slightly better (though still with substantial under-coverage) when $\beta$ is negative. In particular, it can be seen from the comparison between $CV_{AR}$ and $CV_{PZ}$ that $CV_{AR}$ has better coverage performance even though the data-generating mechanism is not an autoregressive process. 

A drawback of AM-PW here is that the bandwidth it uses is based on an $AR(1)$ model for the prewhitened data. As is pointed out by \cite{haan1996inferences}, it is not true in general that the data-dependent bandwidth parameter should solely depend on the first-order autocorrelation of the prewhitened data. The bandwidth selection procedure of AM-PW follows the proposal of \cite{andrews1991heteroskedasticity} and works well when the prewhitened process has a monotonically decreasing spectral density (See \cite{haan1996inferences}). Since this monotonicity may not hold in practice, such a predetermined fitting of an AR(1) model to the prewhitened data may have drawbacks, as seen here. Note that $CV_C$ avoids the use of pilot estimates as it is based on cross-validation. Overall, for the simulations in this subsection, NW-PW outperforms AM-PW, but $CV_C$ is superior. 

Finally, the gap between the observed coverage rate and the nominal coverage rate of $CV_C$  narrows as $n$ goes from 50 to 200.
\newpage
\subsection{$AR(2)$ Processes}
$$X_t = \frac{1}{2}\phi X_{t-1} + \frac{1}{2}\phi X_{t-2} +\varepsilon_t, \qquad \varepsilon_t \stackrel{i.i.d}{\sim} N (0,1)$$

\begingroup
\renewcommand\arraystretch{0.5}
\begin{longtable}[c]{@{}llcccccc@{}}
                     &        & \multicolumn{3}{c}{n = 50} & \multicolumn{3}{c}{n = 200} \\* \midrule
\endfirsthead
\endhead
\bottomrule
\endfoot
\endlastfoot
\textbf{$\phi$}                             & \textbf{Method} & \textbf{90\%} & \textbf{95\%} & \textbf{99\%} & \textbf{90\%} & \textbf{95\%} & \textbf{99\%} \\* \midrule
\multicolumn{1}{c}{\multirow{5}{*}{0.3}} & $CV_C$             & 80.3          & 87.5          & 94.2          & 84.6          & 90.7          & 96.8          \\
\multicolumn{1}{c}{} & $CV_{AR}$ & 80.2    & 87.1    & 94.0   & 85.6     & 91.3    & 97.2    \\
\multicolumn{1}{c}{} & $CV_{PZ}$   & 79.4    & 94.2    & 94.4   & 80.4     & 87.2    & 95.3    \\
\multicolumn{1}{c}{} & AM-PW  & 81.1    & 87.2    & 94.7   & 83.1     & 89.8    & 96.8    \\
\multicolumn{1}{c}{} & NW-PW  & 81.2    & 87.4    & 94.7   & 85.5     & 91.5    & 97.7    \\* \midrule
\multirow{5}{*}{0.5} & $CV_C$    & 75.8    & 82.5    & 90.5   & 84.4     & 90.5    & 96.2    \\
                     & $CV_{AR}$ & 76.4    & 82.8    & 90.6   & 85.7     & 91.2    & 96.7    \\
                     & $CV_{PZ}$   & 72.5    & 80.2    & 89.9   & 77.4     & 84.0    & 92.3    \\
                     & AM-PW  & 74.0    & 81.3    & 90.0   & 77.5     & 84.5    & 93.3    \\
                     & NW-PW  & 75.9    & 83.1    & 91.5   & 82.7     & 89.1    & 96.3    \\* \midrule
\multirow{5}{*}{0.7} & $CV_C$    & 71.4    & 77.7    & 86.8   & 84.5     & 89.8    & 95.8    \\
                     & $CV_{AR}$ & 74.1    & 79.6    & 87.4   & 86.3     & 91.1    & 96.4    \\
                     & $CV_{PZ}$   & 61.9    & 70.3    & 81.9   & 77.4     & 83.9    & 92.7    \\
                     & AM-PW  & 62.5    & 70.6    & 82.3   & 69.2     & 77.9    & 87.9    \\
                     & NW-PW  & 66.9    & 74.0    & 85.1   & 77.0     & 83.8    & 92.8    \\* \midrule
\multirow{5}{*}{0.9} & $CV_C$    & 65.5    & 71.4    & 80.2   & 83.4     & 88.4    & 93.4    \\
                     & $CV_{AR}$ & 68.3    & 74.4    & 82.5   & 84.3     & 89.2    & 94.4    \\
                     & $CV_{PZ}$   & 38.8    & 44.9    & 57.5   & 59.6     & 67.6    & 79.0    \\
                     & AM-PW  & 42.0    & 48.8    & 60.5   & 53.5     & 61.3    & 73.2    \\
                     & NW-PW  & 45.8    & 53.0    & 65.1   & 58.5     & 66.3    & 77.9    \\* \bottomrule
\caption{Coverage Rates for $AR(2)$ Processes}
\label{table10}\\
\end{longtable}
\endgroup

This set of experiments was proposed by \cite{den199712}. The overall best method in this experiment is $CV_C$. The inclusion of parametric autoregressive model candidates in FDCV is motivated by $AR(1)$ prewhitening in the HAC literature. The prewhitening filter that AM-PW and NW-PW considered is a fixed first-order filter, where the method being discussed allows for model selection in choosing a parametric model. The advantage of such flexibility is not clear in experiment 5.1, where the data-generating mechanism is an $AR(1)$ which is exactly the process that AM-PW and NW-PW are designed for. The value of the autoregressive coefficients is the same, taken to be $\frac{1}{2}\phi$. When $\phi$ increases, the autoregressive polynomial will have a root that is close to the unit circle, and the spectral density will be sharper. Notice that in all cases, the actual coverage probabilities are smaller than the nominal coverage probabilities, which is similar to was found in experiment 5.1. The performances of $CV_C$, AM-PW and NW-PW are similar when $\phi= 0.3, 0.5$ for both $n = 50$ and $200$. When $\phi = 0.7$ and $0.9$, AM-PW and NW-PW are strongly outperformed by $CV_C$. This shows the advantage of flexibility in selecting the parametric components for the HAC problem. The inclusion of an $AR(2)$ model candidate in class $\textbf{C}$ is helpful for $CV_C$ in the current situation.

In addition, as observed in 5.1, when $\phi_1 = 0.95$, the use of REML to estimate the autoregressive model improves the performance when the spectral density has a sharp peak in the simulation. In fact, in the earlier stage of this research, the Yule-Walker or Burg method was used to estimate the autoregressive model. It was discovered that using REML significantly improve the performance.

Finally, the gap between the observed coverage rate and the nominal coverage rate of $CV_C$  narrows as $n$ goes from 50 to 200.

\subsection{ $c$ in the Criterion Function}
In this subsection, the data generating process considered is the AR(1) process, $X_t = 0.9X_{t-1} + \varepsilon_t$, with $\varepsilon_t \stackrel{i.i.d}{\sim} N(0,1)$. In this scenario, different values of the exponent $c$ are experimented with, specifically 0.2, 0.5, 0.8, and 0.9. The coverage rates of $CV_C$, $CV_{AR}$, and $CV_{PZ}$ are reported at nominal rates of $90\%$, $95\%$, and $99\%$, based on 3,000 replications with a sample size of 50.

\begin{table}[h]
    \centering

    \label{table_label}
    \begin{tabular}{cccccc}
        \toprule
        \textbf{c} & \textbf{Method} & \textbf{90\%} & \textbf{95\%} & \textbf{99\%} \\
        \midrule
            & $CV_C$   & 59.9\% & 65.4\% & 74.5\% \\
        0.2 & $CV_{AR}$  & 75.7\% & 80.6\% & 86.5\% \\
            & $CV_{PZ}$  & 44.6\% & 51.3\% & 64.3\% \\
        \midrule
            & $CV_C$   & 65.7\% & 71.8\% & 79.2\% \\
        0.5 & $CV_{AR}$  & 75.4\% & 81.3\% & 88.5\% \\
            & $CV_{PZ}$  & 41.5\% & 47.6\% & 58.1\% \\
        \midrule
            & $CV_C$   & 70.8\% & 77.2\% & 84.7\% \\
        0.8 & $CV_{AR}$  & 75.7\% & 81.5\% & 88.0\% \\
            & $CV_{PZ}$  & 46.6\% & 53.6\% & 66.7\% \\
        \midrule
            & $CV_C$   & 71.6\% & 77.8\% & 85.2\% \\
        0.9 & $CV_{AR}$  & 77.2\% & 82.7\% & 89.4\% \\
            & $CV_{PZ}$  & 47.0\% & 54.2\% & 67.5\% \\
        \bottomrule
    \end{tabular}
        \caption{Coverage Rate for an AR(1) process with different values of $c$}
\end{table}

   A significant performance improvement can be observed when changing $c$ from 0.2 to 0.5, and from 0.5 to 0.8. However, when changing $c$ from 0.8 to 0.9, the improvement is minor, even though the computational cost is much higher. The difference in performance becomes even smaller when using a larger sample size instead of 50. In an earlier phase of research, various values of $c$ were tried for other processes, but this scenario can help readers better understand why $c=0.8$ is a reasonable choice.

\section{Conclusion}
\label{sec:conc}
A unified frequency domain cross-validation method is applied to select an estimate of the spectral density at zero frequency, and the performance of confidence intervals for the mean based on the resulting HAC standard error is studied. Unlike classical HAC methods, this method conducts unified model/tuning parameter selection where candidates span across parametric and nonparametric estimators. Specifically, the model/tuning parameters are proposed to be automatically selected from a class of \textbf{C} consisting of REML-based autoregressive spectrum estimators of order 0 to 5 and lag-weights spectrum estimators with Parzen kernel from truncation point 1 to $m(n)$.

The performance of the confidence interval for the proposed data-driven method was studied and compared to other popular plug-in-based approaches like those by \cite{newey1994automatic} and \cite{andrews1992improved} in the case of the mean. It is found that the proposed method is the best performing and the most reliable method in simulation. More specifically, the proposed method has superior performance when the time series has an autoregressive root that is closed to the unit circle due to the inclusion of the REML-based autoregressive estimators. Moreover, the inclusion of autoregressive spectrum estimates can be advantageous even if the time series is not an autoregressive process, such as a moving average process. Finally, the method is reliable in the case of white noise where the spectral density is constant, due to the inclusion of the nonparametric lag-weights estimators and better choice of the bandwidth parameter.

For future work, the consideration of tapering for nonparametric estimators could also be explored. This is motivated by the simulation study where it is found in most cases except for white noise that even if the spectral density around zero is relatively flat, the FDCV would still more frequently select a parametric estimator. Moreover, it was generally found that $CV_{AR}$ has better performance than $CV_{PZ}$ even if the process is not autoregressive. Thus, it is desirable to improve the performance of the nonparametric spectrum estimators. One related work of this topic on the HAC problem is \cite{smith2005automatic} who proposes using multitapering for the HAC standard error estimation. However, his ideas have not yet been verified in simulation so far as available information suggested. In general, tapering the data will reduce the bias at the cost of inflating the variance. However, if smoothing is applied directly on the non-tapered data, the bias will persist as smoothing only dampens the variance. So, to apply a nonparametric estimator on non-tapered data is to smooth out what has already been biased. The goal is to find the tapers that can noticeably reduce the bias without inflating too much of the variance and then applying the nonparametric estimators on the tapered data. 

Although this paper primarily focuses on the location model for simplicity, a natural question arises as to how the method could be extended to the setting
of a general time series regression. This setting is allowed for in the existing HAC literature. To explain this setting, the discussion in \cite{lazarus2018har} is followed. The time series regression model can be written as
\[
Y_t = \beta^{\prime} X_t + \varepsilon_t
\]
where $X_t = (1,x_t^\prime)^\prime$, $\{x_t\}$ is a $q \times 1$ stationary time series with zero mean, $\beta$ is a $((q+1)\times 1)$ vector of regression parameters (starting with an intercept) and $\{\varepsilon_t\}$ is a scalar stationary process with zero mean. Under the assumption that
$E[\varepsilon_t | X_t] = 0$, the HAC problem arises if
$z_t = X_t \varepsilon_t$ are autocorrelated. The key step in constructing HAC standard errors is the estimation of the long-run variance, which is $2 \pi$ times the spectral density of $\{z_t\}$ at zero frequency. Since $\{z_t\}$ is $((q+1)\times 1)$, the long-run variance is a $((q+1)\times(q+1))$ matrix. Note that $z_t$ is not observable, but the usual proxy used for estimation is $\hat z_t = X_t \hat \varepsilon_t$ where $\hat \varepsilon_t$ are the ordinary least squares residuals.

Thus, to extend the method to the time series regression setting it is necessary to consider frequency domain cross validation and spectral estimation for multiple time series. Lag-weights estimation extends in a straightforward manner to the case where the observations are a vector time series, and has been used in this setting in much of the existing
HAC literature. Weighted least squares approximate REML estimators for vector autoregressions are developed in \cite{chen2010weighted}. The definition of the leave-one-out version of a vector time series data set would take the same form as the definition given here, except that now $x_t$ and $J_j$ are vectors. As to the cross-validation function, keeping in mind that in the current setting the periodogram is a $((q+1)\times(q+1))$ matrix, one could work with a generalization of $CV(\hat f)$ based on a matrix logarithm, or consider a generalization of $CVLL (\hat f)$ based on the multiple time series version of the Whittle likelihood (see \cite{dunsmuir1979central}, Equation (1.5)). The implementation of cross validation
methods in the time series regression setting is left for future work.
Furthermore, it is believed that the FDCV method could be quite beneficial for other univariate applications, such as stationarity tests (e.g., \cite{kwiatkowski1992testing}) and unit root tests (e.g., \cite{phillips1988testing}, Exploring its potential in these areas is a direction for future research.

Finally, HAC focuses on robust standard error estimation in a short memory process. In most of the HAC literature, it is assume that the spectral density at zero frequency is finite and positive. However, if the time series has $f(0)$ being infinite or zero (so that the time series has long memory or is anti-persistent), then the HAC methods based on estimating $f(0)$ are no longer consistent. Details of such problems can be found in \cite{robinson2005robust} and \cite{abadir2009two} where an alternative MAC (memory autocorrelation consistent) estimator is considered. The possibility of applying FDCV to the MAC problem is an area worth investigating in future research.

\section*{Acknowledgments}
The authors sincerely thank the Associate Editor and Referees for comments that substantially
improved the quality of the paper.

\bibliographystyle{chicago}
\bibliography{Bibliography-MM-MC}





\end{document}